\documentclass[journal]{IEEEtran}

\ifCLASSINFOpdf
\else
\fi

\usepackage{setspace}
\usepackage{bbm}
\usepackage[cmex10]{amsmath}
\usepackage{amssymb}
\usepackage{cite}
\usepackage{graphicx}
\usepackage{array,color}
\usepackage{amsmath}
\allowdisplaybreaks
\usepackage{stfloats}
\usepackage{graphicx}
\usepackage{epstopdf}
\usepackage{subfigure}
\usepackage{tabularx}
\usepackage{epsfig,epsf,color,balance,cite}
\usepackage{algorithmic}
\usepackage{algorithm}
\usepackage{url}
\usepackage{bm}
\usepackage{footnote}
\usepackage{amsthm}
\newcounter{example}[section]

\newtheorem{theorem}{Theorem}
\newtheorem{lemma}[theorem]{Lemma}
\newtheorem{proposition}[theorem]{Proposition}

\ifodd 1
\usepackage{soul}
\usepackage{color}
\setstcolor{red}

\newcommand{\del}[1]{\st{#1}} 

\newcommand{\com}[1]{\textbf{\color{red} (COMMENT: #1)}} 
\newcommand{\response}[1]{\textbf{\color{green} (RESPONSE: #1)}} 
\else

\newcommand{\del}[1]{}
\allowdisplaybreaks[4]
\newcommand{\com}[1]{}
\newcommand{\comg}[1]{}
\newcommand{\response}[1]{}
\fi
\allowdisplaybreaks[4]

\title{\vspace{-0.5cm}{\LARGE Intelligent Reflecting Surface-Aided Spectrum Sensing for Cognitive Radio}\vspace{-0.2cm}}

\author{Shaoe~Lin, Beixiong~Zheng, {\it Member, IEEE}, Fangjiong~Chen, {\it Member, IEEE}, and~Rui~Zhang, {\it Fellow, IEEE}
\vspace{-1.0cm}

\thanks{This work was supported by National Key Research and Development Program of China under Grant 2020YFB1807700, by Guangdong Provincial Key Laboratory of  Short-Range Wireless Detection and Communication under Grant  2017B030314003, by Ministry of Education, Singapore under Award T2EP50120-0024, by Advanced Research and Technology Innovation Centre (ARTIC) of National University of Singapore under Research Grant R-261-518-005-720, and by China Scholarship Council.  (\emph{Corresponding author: Beixiong Zheng.}) }

	\thanks{
		S. Lin and F. Chen are with the School of Electronic and Information Engineering, South China University of Technology, Guangzhou 510641, China (e-mails: eeshe.lin@mail.scut.edu.cn, eefjchen@scut.edu.cn).
		
		B. Zheng and R. Zhang are with the Department of Electrical and Computer Engineering, National University of Singapore,  Singapore 609774 (e-mails: elezbe@nus.edu.sg, elezhang@nus.edu.sg).
}
}

\begin{document}
\markboth{IEEE Wireless Communications Letters, Vol. XX, No. XX, February 2022}{Lin}
\maketitle

\begin{abstract}

Spectrum sensing is a key enabling technique for cognitive radio (CR), which provides essential information on the spectrum availability. 
However, due to severe wireless channel fading and path loss, the primary user (PU)  signals received at the CR or  secondary user (SU) can be practically too weak for reliable detection. 
To tackle this issue, we consider in this letter a new intelligent reflecting surface (IRS)-aided spectrum sensing scheme for CR, by exploiting the large aperture  and passive beamforming gains of IRS to boost the PU signal strength received at the SU to facilitate its  spectrum sensing. 
Specifically, by dynamically changing the IRS reflection over time according to a given codebook, its reflected signal power varies substantially at the SU, which is utilized for opportunistic signal detection.
Furthermore, we propose a weighted energy detection method by combining the received signal power values over different IRS reflections, which significantly improves the detection performance. 
Simulation results validate the performance gain of the proposed IRS-aided spectrum sensing scheme, as compared to different benchmark schemes. 

\end{abstract}
\IEEEpeerreviewmaketitle
\vspace{-0.2cm}

\section{Introduction}
\IEEEPARstart{C}{ognitive radio} (CR), which allows the secondary users (SUs) to opportunistically communicate over the  spectrum bands  allocated to a set of primary users (PUs) in a given area, is a revolutionizing technology to enable dynamic spectrum access in wireless communications. 
In particular, spectrum sensing is a key enabling technique for CR to identify the available spectrum for opportunistic spectrum access. 
Specifically, in the spectrum sensing-enabled CR network, the SU first detects/senses the presence of any active PU transmissions over the band of interest,  then decides to  transmit its own message over this band  if the sensing result indicates that it is currently unoccupied by any PUs, thus  improving the bandwidth efficiency of the network.
The appealing function of spectrum sensing has motivated substantial studies on its algorithm design, 
such as energy detection \cite{Ali2017Advances,Sobron2016Energy}, matched-filter (coherent) detection \cite{Zhang2014Matched}, and cyclostationary (feature) detection \cite{Huang2013On}, among others. 
However, due to various wireless channel impairments such as shadowing, multipath fading, and substantial path loss over long distance, the PU signals received at the SU can be very weak in practice, thus resulting in unsatisfactorily high missed-detection probability and/or false-alarm probability, which further degrades the dynamic spectrum access efficiency of CRs/SUs. 
Moreover, as the PU location is usually random and time-varying, its channel to the SU is also random and practically uncontrollable.   

Recently, the technological advance in digitally-controlled and dynamically-tunable  metasurfaces has made it feasible to reshape wireless propagation channels in a cost-effective manner, which leads to the promising new technology of intelligent reflecting surface (IRS)-aided wireless communications  \cite{Wu2021Intelligent,zheng2019intelligent,Beixiong2021A}. 
Specifically, IRS consists of a large number of passive reflecting elements that can be reconfigured by a controller in tuning their reflection amplitudes and/or phase shifts, thus collaboratively boosting/suppressing its reflected signal power in designated directions. 
Hence, IRS provides a new and cost-effective means to compensate wireless channel fading and path loss and make them programmable in real time.
The appealing features of IRS have motivated extensive studies on their applications in various wireless systems, including some primitive works on IRS-aided CR systems for spectrum sharing \cite{Guan2020Joint} or spectrum sensing \cite{Li2020IRS}. 
However, these works assumed that either  instantaneous or statistical channel state information (CSI) on the PU is available to the SU, which may be difficult to be practically achieved due to the lack of cooperation  between the primary and secondary systems.

\begin{figure}[!t]
	\centering
	\includegraphics[width=1.6 in]{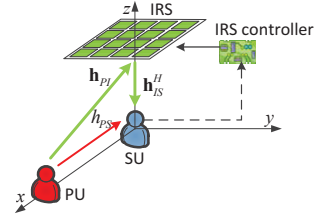}	\vspace*{-0.35cm}
	\caption{{An illustration of IRS-aided spectrum sensing in the CR system. }}
	\label{system_block}
	\vspace*{-0.6cm}
\end{figure}

Motivated by the above, we consider in this letter a new IRS-aided spectrum sensing scheme for CR, where a SU detects the presence of the PU signals aided by the large aperture and passive beamforming gains of the IRS, as shown in Fig.~\ref{system_block}. 
Specifically,  by sequentially changing the IRS reflection over time according to a given codebook, its reflected signal power varies substantially at the SU receiver to facilitate its opportunistic signal detection. 
Furthermore, we devise  a  weighted energy detection (WED) method customized for the proposed  IRS-aided spectrum sensing, which optimizes the weighted coefficients for combining the received signal power values over different IRS reflections to minimize the missed-detection probability at a given false-alarm probability. 
Finally, we present simulation results to validate the performance gain of the proposed IRS-aided spectrum sensing scheme as compared to  various benchmark schemes.

\section{System Model}\label{sm}

We consider the IRS-aided spectrum sensing in a CR system as illustrated in Fig.~\ref{system_block}, 
where a SU performs spectrum sensing based on a frame of $N$ discrete-time observations to detect the presence of any PU signals in a given frequency band\footnote{For convenience, we consider the narrow-band channel for all links in this paper, while the proposed design can be directly applied to general wide-band channels with independent sensing over parallel narrow bands at the SU.}, aided by an IRS that consists of $L$ passive reflecting elements mounted above the SU. 
In practice, to maximize the IRS reflection power, the IRS is desirable to be deployed near either the PU transmitter or the SU receiver\footnote{Alternatively, we may consider  deploying multiple IRSs around the SU receiver in a distributed manner, which can be regarded as deploying a larger-size IRS near the SU receiver equivalently.} (shown in Fig.~\ref{system_block}).   The IRS is connected to a smart controller that is responsible for dynamically adjusting the phase shifts of its passive reflecting elements and synchronizing with the SU via a control link. 
For ease of exposition, we assume that there is only one PU in the sensing range of the SU and both of the PU and SU are equipped with one single antenna. 
The channels from the PU to SU, from the PU to IRS, and from the IRS to SU are denoted by $h_{PS} \in \mathbb{C}$, ${\bf h}_{PI}\in \mathbb{C}^{L\times 1}$, and ${\bf h}_{IS}^H \in \mathbb{C}^{1 \times L}$, respectively. We assume that $h_{PS}$, ${\bf h}_{PI}$, and ${\bf h}_{IS}^H$ remain constant during the detection interval of $N$ samples. 
Note that it  is practically difficult to obtain the CSI of $h_{PS}$ and ${\bf h}_{PI}$ at the SU/IRS due to the lack of information sharing between the PU and SU/IRS. 
Therefore, we consider in this letter a challenging but practical scenario where the instantaneous CSI of the above-mentioned channels is unknown. 
To facilitate the detection of the primary transmission over unknown channels, we consider a codebook-based IRS-aided spectrum sensing scheme, where the IRS dynamically changes its reflection according to a pre-designed codebook during the sensing period. 

Specifically, each detection frame of $N$ observations is divided into $M$ blocks, denoted by $\mathcal{M}\triangleq\{1,\ldots,M\}$, each of which consists of $\bar{N}=N/M$ (assumed to be an integer for convenience) consecutive observations. 
During the interval of each block, a passive reflection state is sequentially drawn from the per-designed codebook and applied at the IRS. 
Let ${\bm \Sigma}_m = \text{diag} \left(  e^{j\theta_{1,m}}, e^{j\theta_{2,m}}, \ldots, e^{j\theta_{L,m}} \right) $ denote the IRS reflection state during the interval of block $m$, where $\theta_{l,m} \in [0,2\pi)$ represents the phase shift of the $l$-th passive reflecting element, with $l=1,2,\ldots,L$ and $m\in\mathcal{M}$. Then, for the case in which the PU is transmitting, the effective channel from the PU to SU during the interval of  block $m$ is given by
\begin{align}\label{effective channel}
	g _m= h_{PS} + {\bf h}_{IS}^H {\bm \Sigma}_m   {\bf h}_{PI} , \quad m=1,2,\ldots,M
\end{align}
which dynamically varies over blocks due to the adjustment on the IRS reflection. 
Therefore, the hypothesis testing problem of our interest for the SU's signal detection is expressed as
\begin{align} \label{hypothesis M} \left\{
	\begin{array}{l}
		\hspace*{-6pt}\mathcal{H}_0:  y_m [i] = n_m [i] \\
		\hspace*{-6pt}\mathcal{H}_1:  y_m [i] =  g _m s_m[i] +n_m[i] 
	\end{array}\right. 
\end{align}
where  $\mathcal{H}_0$ and $\mathcal{H}_1$ denote the absence and presence of the PU signal, respectively, $y_m [i]$ is the $i$-th discrete-time observation in the $m$-th block with $m=1,\ldots,M$ and $i=0,\ldots,\bar{N}-1$, $n_m [i]$ is the noise at the SU, 
modeled as an independent and identically distributed (i.i.d.) circularly symmetric complex Gaussian (CSCG) random variable with zero mean and variance $\sigma^2$, and $s_m [i]$ is the signal transmitted by the PU transmitter with the average power of $P_t$, i.e., $\mathbb{E} \left[ s_m [i] \right]=P_t $.

\section{Detection Method}

In this section, we propose a new WED method for solving the hypothesis testing problem in \eqref{hypothesis M} to leverage IRS-induced channel gain variations for enhancing the sensing performance. 
Specifically, the test statistic is expressed as 
\begin{align}\label{statistic}
	T_{ }\triangleq\sum_{m=1}^{M} w_m T_m \overset{ \mathcal{H}_1 }{ \underset{ \mathcal{H}_0 }{\gtrless} } \lambda
\end{align}
where $\lambda > 0 $ is the detection threshold, 
$w_m$ is the weighted coefficient of block $m$ with $w_m \ge 0$ and $\sum_{m=1}^{M} w_m=1$, 
and $T_m$ is the average received power during block $m$ normalized by the noise variance\footnote{The noise variance $\sigma^2$ is assumed to be known, which can be obtained by applying noise power estimation at the SU receiver before sensing.},  
i.e.,
\begin{align}\label{tm}
	T_m \triangleq \frac{1}{ \bar{N} \sigma^2} \sum_{i=0}^{\bar{N}-1} |  y_m [i] |^2, \quad m=1,2,\ldots,M.
\end{align} 
Note that based on \eqref{hypothesis M}, $\{y_m [i]\}_{i=0}^{\bar{N}-1}$ are  i.i.d. CSCG random variables with the distribution given by  
\begin{align}
	y_m [i] \sim \left\{ 	\begin{array}{ll}
		\mathcal{CN} \left( 0 , \sigma^2 \right) , & \text{under}~ \mathcal{H}_0\\
		\mathcal{CN} \left( g _ms_m [i] , \sigma^2 \right) , & \text{under}~\mathcal{H}_1 .\\
	\end{array} \right.  
\end{align}
Let $X \triangleq 2\bar{N}T_m$; then  according to \eqref{tm}, $X$ can be regarded as the sum of the squares of $2 \bar{N}$ i.i.d. Gaussian random variables each with unit variance. 
Therefore, it can be verified that $X$ follows a central chi-square distribution with $2 \bar{N}$ degrees of freedom under $\mathcal{H}_0$, and a noncentral chi-square distribution with $2 \bar{N}$ degrees of freedom and a noncentrality parameter of $2  \bar{N} \varrho | g _m|^2 $ under $\mathcal{H}_1$, where $	\varrho \triangleq P_t/\sigma^2$. 
As such, by assuming that $\bar{N}$ is sufficiently large and according to the central limit theorem, 
$T_m$ can be well approximated by the following Gaussian distribution, 
\begin{align}\label{distribution Tm appr}
	T_m  \sim \left\{ 	\begin{array}{ll}
		\hspace*{-3pt}\mathcal{N} \left( 1 , \dfrac{1}{ \bar{N}} \right) , & \text{under}~\mathcal{H}_0\\
	\hspace*{-3pt}	\mathcal{N} \left( 1 + \varrho| g _m|^2  ,  \dfrac{1 + 2 \varrho| g _m|^2 }{ \bar{N}}  \right) , & \text{under}~\mathcal{H}_1.\\
	\end{array} \right. 
\end{align}

Next, we design the weighted coefficients $w_m$'s in \eqref{statistic} to minimize the missed-detection probability subject to a given requirement on the maximum false-alarm probability. 
In particular, if $g _m$'s are known {\it a priori}, it has been shown in \cite{LiangYingChang2008Sensing} that the optimal weighted coefficients 
are given by 
\begin{align}\label{optimal coefficients}
w_m^* = \frac{| g_m |^2}{\sum_{k=1}^{M}  | g_k |^2 }, \quad m=1,2,\ldots,M .
\end{align}
Unfortunately, this optimal design is inapplicable to our considered setup due to the lack of prior knowledge of $g _m$'s.  
Nevertheless, the design in \eqref{optimal coefficients} still motivates us to devise  a practical WED scheme by approximating \eqref{optimal coefficients}.

Let us now rewrite \eqref{optimal coefficients} by multiplying both of its numerator and denominator by $	\varrho $ as follows:
\begin{align}\label{optimal coefficients gamma}
	w_m^* = \frac{ \varrho| g _m|^2  }{\sum_{k=1}^{M}  \varrho| g _k|^2  } , \quad m=1,2,\ldots,M 
\end{align}
where $	\varrho| g_m|^2  $ denotes the received signal-to-noise ratio  during block $m$. 
As $ g _m  $'s are assumed to be unknown in our considered setup, 
the exact values of $	\varrho| g _m|^2  $'s are generally unknown as well. 
Nevertheless, by noting  the distributions of $T_m$'s under $\mathcal{H}_1$ given in \eqref{distribution Tm appr} and the inherent non-negativity restrictions on $\varrho| g _m|^2 $'s, we take $[T_m - \alpha ]^+$ as an estimate/approximation of $\varrho| g _m|^2 $ and substitute $\varrho| g _m|^2  \approx [T_m - \alpha ]^+$ into \eqref{optimal coefficients gamma}, which leads to the following weighted coefficients:
\begin{align}\label{practical_combining}
	w_m=\frac{ [T_m - \alpha  ]^+}{\sum_{k=1}^{M}  [ {T_k} - \alpha ]^+ }, \quad m=1,2,\ldots,M
\end{align}
where $\alpha\in[0,1)$ is a scaling factor for controlling the approximation of \eqref{optimal coefficients gamma} and $[x]^+ \triangleq \max(x,0)$. 
Note that those blocks with the average received power less than or equal to $\alpha$, denoted by $\mathcal{J}=\{ m | T_m \le \alpha, m\in \mathcal{M} \}$, are discarded by assigning a zero weighted coefficient according to \eqref{practical_combining}. 

\emph{Remark 1:} An intuitive impact of $\alpha$ on the detection performance  can be envisioned as follows: if $\alpha$ is too small, $[T_m - \alpha ]^+$ under $\mathcal{H}_1$ will deviate from $\varrho| g _m|^2 $ due to the noise effect, which may increase the missed-detection probability; while too large value of $\alpha$ results in fewer effective blocks (due to the increasing number of invalid blocks, $|\mathcal{J}|$) for averaging out the noise effect, which may increase the  false-alarm probability. 
As such, there exists an interesting trade-off in selecting the proper value of $\alpha$ to balance the performances of  false-alarm and missed-detection probabilities, as will be shown via simulation results  in Section \ref{sim}.

\section{Performance Analysis}

In this section, we derive a closed-from expression for the  false-alarm probability of the proposed WED scheme, based on which the detection threshold can be uniquely determined. Then, we analyze the performance upper bound (UB) on the missed-detection probability of the proposed scheme.  

\subsection{False-Alarm Probability}

First, we derive a closed-form expression for the false-alarm probability of the proposed scheme, defined as
$
P_{\rm FA}^{ } (\lambda) \triangleq {\rm Pr} \left(T_{ } > \lambda | \mathcal{H}_0 \right) 
$. 
By substituting \eqref{practical_combining} into \eqref{statistic} and relaxing the non-negativity restrictions, $
P_{\rm FA}^{ } (\lambda) $  can be expressed as
\begin{align}
	P_{\rm FA}^{ }  (\lambda) = &{\rm Pr} \left( \left. \frac{ \sum_{m=1}^{M} (T_{{m}} - \alpha   ) T_{{m}} }{\sum_{ k=1}^{M}  ( T_{k} - \alpha  ) } > \lambda \right| \mathcal{H}_0 \right)  
\end{align}
which can be further simplified as
\begin{align}
\hspace*{-3pt}	P_{\rm FA}^{ } (\lambda)
	\hspace*{-1pt}=\hspace*{-1pt} {\rm Pr} \left( \left. \hspace*{-1pt}{ \sum_{ m=1}^{M} \hspace*{-2pt} \left( T_{{m}}^2 \hspace*{-1pt} - \hspace*{-1pt} ( \alpha    + \lambda ) T_{{m}} \right) } \hspace*{-2pt}> \hspace*{-2pt} - \lambda \alpha  { M}  \right| \mathcal{H}_0 \hspace*{-1pt} \right) \label{Pf 0} .
\end{align}
By letting $c = {(\alpha   + \lambda)}/{2} $, 
we obtain
\begin{align} 
	P_{\rm FA}^{ }  (\lambda) = {\rm Pr} \left( \left. { \sum_{m=1}^{M} ( T_{{m}} - {c} )^2 } > \frac{M}{4} (\alpha  - \lambda)^2 \right| \mathcal{H}_0 \right) \label{Pf} . 
\end{align}
Let $Y \hspace*{-3pt} \triangleq \hspace*{-3pt} {  \bar{N} } \sum_{ m=1}^{M} ( T_{{m}} - c )^2 $; then  based on \eqref{distribution Tm appr}, it can be verified that $Y$ under $\mathcal{H}_0$ follows a noncentral chi-square distribution with $ M$ degrees of freedom and a noncentrality parameter $\underline{\mu}_0 \hspace*{-2pt} = \hspace*{-2pt} { ( 1 - c )^2  }{   \bar{N} }M$. 
We assume that $M$ is sufficiently large\footnote{This assumption is made merely for deriving a simpler form of the false-alarm probability expression, while the proposed spectrum sensing scheme also works when $M$ is finite, as will be shown via simulation results.}; then by using the central limit theorem again, $Y$ can be approximated by a Gaussian random variable with the mean of $M+\underline{\mu}_0$ and the variance of $2 M+ 4 \underline{\mu}_0$. 
Thus, $P_{\rm FA} (\lambda)$ in \eqref{Pf} can be obtained as
\begin{align}
	P_{\rm FA}^{ } (\lambda) = Q \left( \frac{ { ( 1 - \alpha ) \left( {\lambda}  -1 \right)  \bar{N} { M} - { M} }}{ \sqrt{ 2 {M}+ \left( 2- \alpha - {\lambda} \right)^2  \bar{N} {M} } } \right) \label{Pf 1} 
\end{align}
where $Q (x) = \frac{1}{ \sqrt{2 \pi} } \int_x^\infty \exp \left( -\frac{u^2}{2} \right) du $ is the tail probability of a standard Gaussian random variable. 
Accordingly, for a given  false-alarm probability, denoted by ${ P}_{\rm FA}^\star$, the detection threshold $\lambda$ can be uniquely determined as follows:
\begin{align}
	\lambda=  \frac{ ( 1 - \alpha ) \left(  {M} -\varLambda^2 \right) \hspace*{-2pt} + \hspace*{-2pt} \varLambda \sqrt{(1-\alpha)^4  \bar{N} {M} \hspace*{-2pt} + \hspace*{-2pt} \dfrac{ {M}-2 \varLambda^2}{ \bar{N}}} }{ (1-\alpha)^2  \bar{N}  {M} - \varLambda^2 } \hspace*{-2pt} + \hspace*{-2pt} 1 \label{threshold}
\end{align}
where $\varLambda=Q^{-1} \left( {P}_{\rm FA}^\star \right) $. 

\subsection{Missed-Detection Probability}\label{analysis pd}

Next, we characterize the UB on the missed-detection probability of the proposed WED scheme. 
We consider the Rayleigh fading channel model for the PU-IRS link, i.e.,  
${\bf h}_{PI} \sim \mathcal{CN} (\bm 0, \beta_{PI} \bm I )$, for the worst case without any obvious LoS channel component in the link. 
On the other hand, due to the practically short  propagation distance between the IRS and SU, we assume that ${\bf h}_{IS}^H $ is an LoS dominant channel with the large-scale path loss denoted by $\beta_{IS}$.\footnote{The analysis for  the case of PU-side IRS is similar to that for the considered case of SU-side IRS by swapping the channel models of the PU-IRS and IRS-SU links.}

\begin{lemma}
With $L\to \infty$, the effective channel gains $| g _m|^2 $'s can be well approximated by i.i.d. noncentral chi-square distributed random variables, i.e.,
\begin{align}\label{gm distribution}
| g _m|^2  \sim \frac{ L \beta_{PI} \beta_{IS}  }{2}   \chi_{2}^2 \left( \frac{2 | h_{PS} |^2}{ L \beta_{PI} \beta_{IS}  }  \right),   m=1,2, \ldots, M. 
\end{align}

\end{lemma}

\begin{IEEEproof}
	The proof is somewhat similar to that in \cite[Lemma 1]{Qin2021Intelligent} and thus is omitted for brevity. 
\end{IEEEproof}

\begin{proposition}\label{proposition 0}
	The missed-detection probability of the proposed WED scheme, defined as $P_{\rm MD}^{ }  \triangleq 1-{\rm Pr} \left(T_{ } > \lambda | \mathcal{H}_1 \right) $, is upper-bounded by
	\begin{align}
		{P}_{\rm MD}^{\rm  }  \hspace*{-2pt}\le \hspace*{-2pt} Q \hspace*{-2pt} \left( \hspace*{-2pt} \frac{ 1 \hspace*{-2pt} + \hspace*{-2pt} \varrho \left(L \beta_{PI} \beta_{IS}  + | h_{PS} |^2\right) \hspace*{-2pt} - \hspace*{-2pt} \lambda }{ \sqrt{ 1 + 2\varrho \left( L \beta_{PI} \beta_{IS}  + | h_{PS} |^2 \right)  } } \hspace*{-2pt} \sqrt{ M  \bar{N} } \hspace*{-2pt} \right)  \hspace*{-3pt} \triangleq  \hspace*{-2pt} \bar{P}_{\rm MD}^{\rm  } (\lambda).  \label{pd lb}
	\end{align}
\end{proposition}

\begin{IEEEproof}
	See Appendix. 
\end{IEEEproof}

Based on $\bar{P}_{\rm MD}^{\rm  }  (\lambda)$ in \eqref{pd lb}, we can further obtain a lower bound on the mean value of the test statistic $T$ under $\mathcal{H}_1$, i.e., 
\begin{align}
	\mathbb{E} \left[ T|\mathcal{H}_1 \right] \hspace*{-2pt} \ge \hspace*{-2pt} \int_{0}^{\infty} \hspace*{-6pt} \lambda d \bar{P}_{\rm MD}^{\rm  }  (\lambda) \hspace*{-2pt}=\hspace*{-1pt}
	1 \hspace*{-2pt}+\hspace*{-2pt}  { \varrho | h_{PS} |^2 \hspace*{-2pt}+\hspace*{-2pt} \varrho L \beta_{PI} \beta_{IS}   }   \label{mean h1 lb} .
\end{align}
From the above, we see that owing to the IRS aperture gain for achieving more signal reflection power at the SU, the considered IRS-aided spectrum sensing scheme has an effective gain of $ \varrho L \beta_{PI} \beta_{IS} $ over the conventional system without IRS, which can be regarded as a special case of the considered system with $L=0$, in terms of the mean value of the test statistic. 
On the other hand, when $ M \to \infty $, we can compute the mean value of $T$ under $\mathcal{H}_0$ with $\alpha=0$ as
\begin{align}\notag
	\mathbb{E} \left[ T|\mathcal{H}_0 \right] \hspace*{-2pt} \approx \hspace*{-2pt} \frac{  \mathbb{E} \hspace*{-2pt} \left[ \hspace*{-2pt} \left(T_{m} - \mathbb{E} \hspace*{-2pt} \left[ T_m |\mathcal{H}_0  \right]  + \mathbb{E} \hspace*{-2pt} \left[ T_m |\mathcal{H}_0 \right]  \right)^2 |\mathcal{H}_0  \right] }{  \mathbb{E} \left[ T_m |\mathcal{H}_0 \right] } \hspace*{-2pt} = \hspace*{-2pt} \frac{1}{\bar{N}} \hspace*{-2pt} + \hspace*{-2pt} 1  
\end{align}
according to \eqref{distribution Tm appr}. 
Let $\Delta$ denote the gap between $\mathbb{E} \left[ T|\mathcal{H}_1 \right] $  and $\mathbb{E} \left[ T|\mathcal{H}_0 \right] $. With $ \bar{N} \to \infty $, we have
\begin{align}
	\Delta \triangleq \mathbb{E} \left[ T|\mathcal{H}_1 \right] - \mathbb{E} \left[ T|\mathcal{H}_0 \right] \ge {\varrho \left( L \beta_{PI} \beta_{IS}  + | h_{PS} |^2 \right) } .
\end{align}
This implies that by increasing the number of passive reflecting elements $L$, the gap between the mean values of $T$ under $\mathcal{H}_0$ and $\mathcal{H}_1$ can be increased, which facilitates to detect the existence of the PU signal from noise.

\section{Numerical Results}\label{sim}
\begin{figure}[!t]  
	\centering    
	\hspace{-0.25cm} \subfigure [$P_{\rm FA}$ and $P_{\rm MD}$ versus $\lambda$]{
		\label{fap&mdp}     
		\includegraphics[width=1.65in]{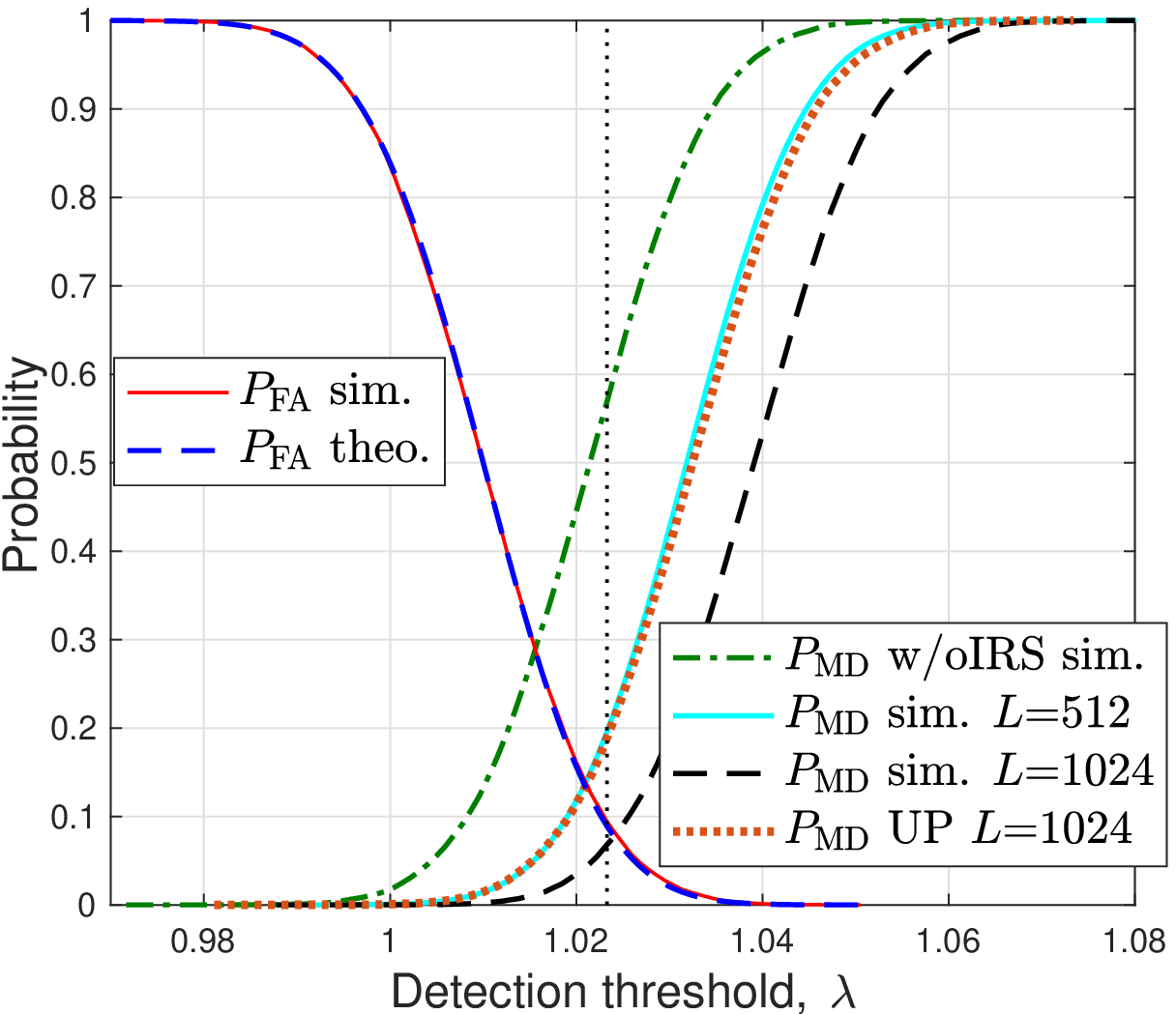} 
	}    
	\centering    
	\hspace{0.00cm}\subfigure [Probability density function of $T$]{
		\label{pdf}     
		\includegraphics[width=1.65in]{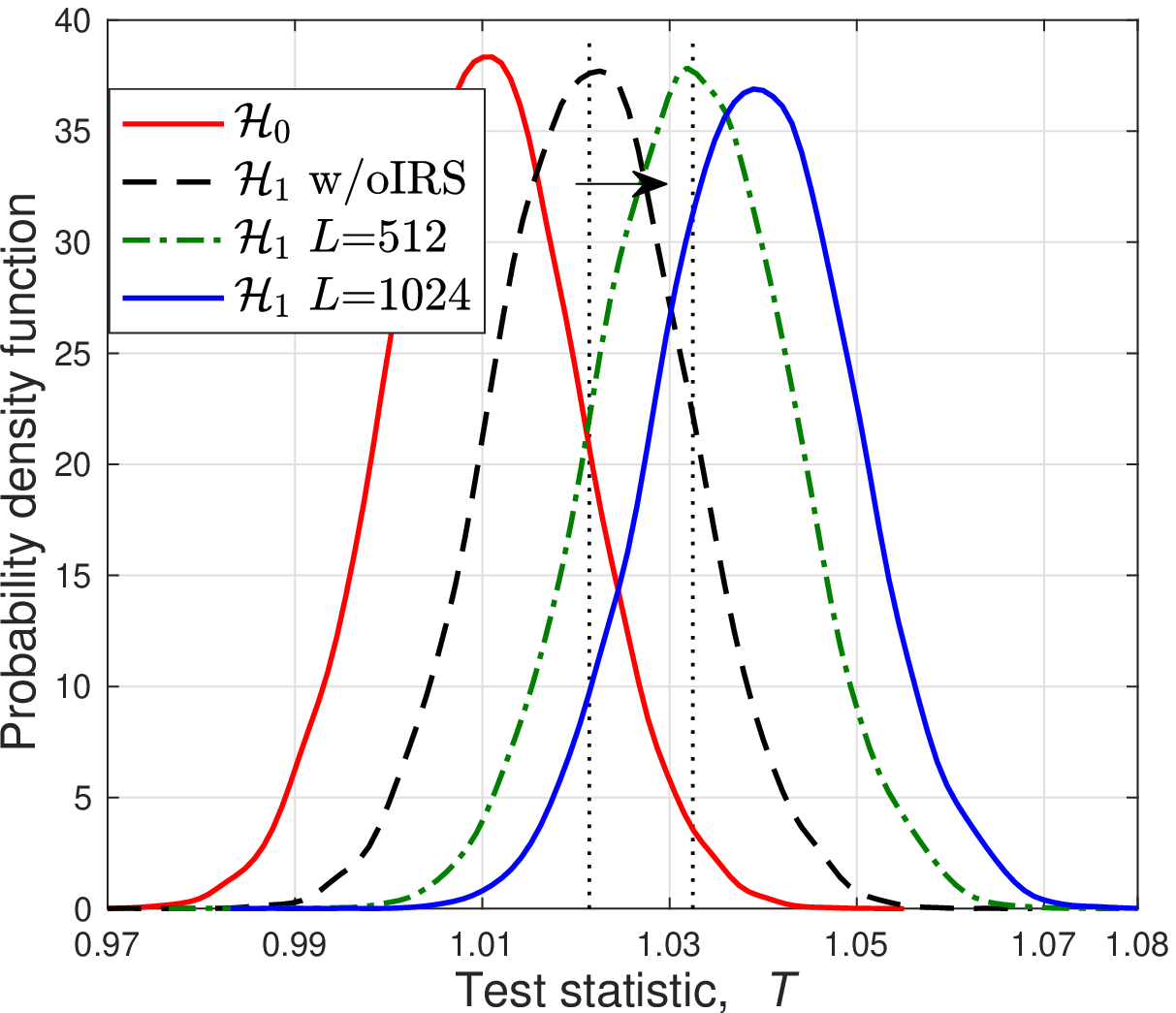} 
	}            \vspace*{-0.4cm}
	\caption{Performance comparison of WED scheme with versus  without IRS, with $M=100$, $ \bar{N}=100$, $\alpha=0$, and $P_t =6$~dBm. }     
	\label{fig1:w&w/oIRS}   
	\vspace*{-0.5cm}
\end{figure}

In this section, we present numerical results to  validate the effectiveness of the proposed WED scheme for IRS-aided spectrum sensing. 
As shown in Fig.~\ref{system_block}, we consider a three-dimensional coordinate system where the SU, PU, and central element of the IRS are located at $(0,0,0)$, $(d \sin \phi, d \cos \phi,0)$, and $(0, 0, 1)$ in meter (m),  respectively, where $\phi$ is the PU's azimuth angle.  
The SU's sensing range  is set to  $R=80$ m. 
We consider the worst case where the PU is located at the edge of the SU's sensing coverage area, i.e., $d=R$ and $\phi$ is uniformly and randomly distributed in $ [0,2\pi)$. 
The IRS is equipped with a uniform rectangular array with half-wavelength spacing, which is located parallel to the $x$-$y$ plane. 
We consider the LoS channel model for the IRS-SU link and the Rayleigh fading channel model for the PU-SU and PU-IRS links, in accordance with Section \ref{analysis pd}. 
The path loss exponents of the IRS-SU, PU-SU, and PU-IRS links  are set as 2, 3.5, and 3.5, respectively, and the reference path loss at a distance of 1 m is set as $30$ dB for all individual links. 
Moreover, we consider the random IRS reflection for generating the codebook, in which each codeword ${\bm \Sigma}_m$ is independently generated with random phase shifts $\theta_{l,m}$'s following the uniform distribution within $ [0,2\pi)$. 
The noise power at the SU is set as $\sigma^2=-70$~dBm and the number of observations within one detection frame is set as $N=10^4$. 
The simulation results are averaged over 1,000 channel realizations.

\begin{figure}[!t] 
	\centering    
	\includegraphics[width=1.65in]{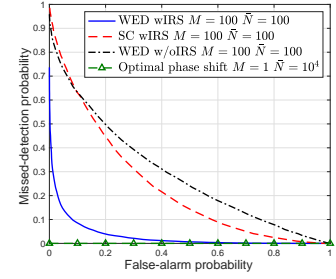} \vspace*{-0.3cm}
	\caption{ROC curves of various combining schemes, with $P_t =6$~dBm and $L=1024$. }     
	\label{roc}   
	\vspace*{-0.6cm}
\end{figure}

The false-alarm and missed-detection probabilities of the proposed scheme are plotted versus the threshold $\lambda$ in Fig.~\ref{fap&mdp}, with $M=100$, $ \bar{N}=100$, $\alpha=0$, and $P_t =6$~dBm. 
The missed-detection probability achieved by the conventional system without IRS is plotted for comparison. 
The theoretical UB on the missed-detection probability of the proposed scheme given in \eqref{pd lb} is also plotted with $L=1024$. 
It is observed that the theoretical analysis of the false-alarm probability given in \eqref{Pf 1} is in perfect agreement with the simulation result. 
Moreover, given a false-alarm probability, e.g., $P_{\rm FA}=0.1$ (or equivalently, $\lambda=1.023$), the proposed scheme aided by IRS achieves a much lower missed-detection probability as compared to the scheme without IRS. 
This can be explained  by comparing the probability distributions of the test statistic $T$ under $\mathcal{H}_1$ with versus  without IRS in Fig.~\ref{pdf}. 
Specifically, we show  in Fig.~\ref{pdf} the probability density function (PDF) of the test statistic $T$, under the same system setups as in Fig.~\ref{fap&mdp}. 
It can be observed that owing to the dynamic IRS reflection for opportunistically reflecting more signal power towards the SU, the PDF of $T$ under $\mathcal{H}_1$ with IRS shifts to the right as compared to that without IRS, thus dramatically reducing its ``overlap area" with the PDF of $T$ under $\mathcal{H}_0$ and achieving a much lower missed-detection probability. 
Moreover, by increasing the number of passive reflecting elements $L$, the missed-detection probability of the proposed sensing system decreases, which is expected due to the larger IRS aperture gain for signal reflection.

In Fig.~\ref{roc}, we show the receiver operating characteristics (ROC) curve of the proposed WED scheme, which constitutes all the achievable pairs of false-alarm and missed-detection probabilities, with $L=1024$, $M=100$, $\bar N=100$, $\alpha=0.2$, and $P_t =6$~dBm. 
The following benchmark schemes are considered for   comparison: 
1) {\bf selection combining (SC) scheme} where the weighted coefficients are given by $g_k^\text{SC}=1$ with ${k}=\arg \max_{m\in \mathcal{M}}  T_m$ and $g_m^\text{SC}=0$ for  $\forall m\in \mathcal{M}\setminus \{k\}$; 
2) {\bf WED scheme without IRS};   
3) {\bf optimal phase shift scheme} with $M=1$, $\bar N=10^4$, and the IRS phase shifts given by 
$
	\theta_{l,1}^*=\angle h_{PS}-\angle [{\bf h}_{IS}^H]_l -\angle [{\bf h}_{PI}]_l$ for $ l=1,\ldots,L 
$. 
It is observed that the optimal phase shift scheme achieves the best performance since the PU signal power received at the SU is maximized by aligning the direct and IRS-reflected channels based on the perfect instantaneous CSI. 
For the practical case without CSI, the proposed WED scheme with IRS significantly outperforms the SC benchmark scheme. 
In particular, the SC scheme performs even worse than the WED scheme without IRS in the low false-alarm probability region. 
This is due to the fact that the number of selected observations in the SC scheme is not sufficient to average out the noise effect, which inevitably leads to a much larger variance in the PDF of its test statistics and thus suffers from a high false-alarm probability.

\begin{figure}[!t] 
	\centering    
	\hspace{-0.25cm} \subfigure [Missed-detection probability versus $M$ ($\alpha=0.5$)]{
		\includegraphics[width=1.65 in]{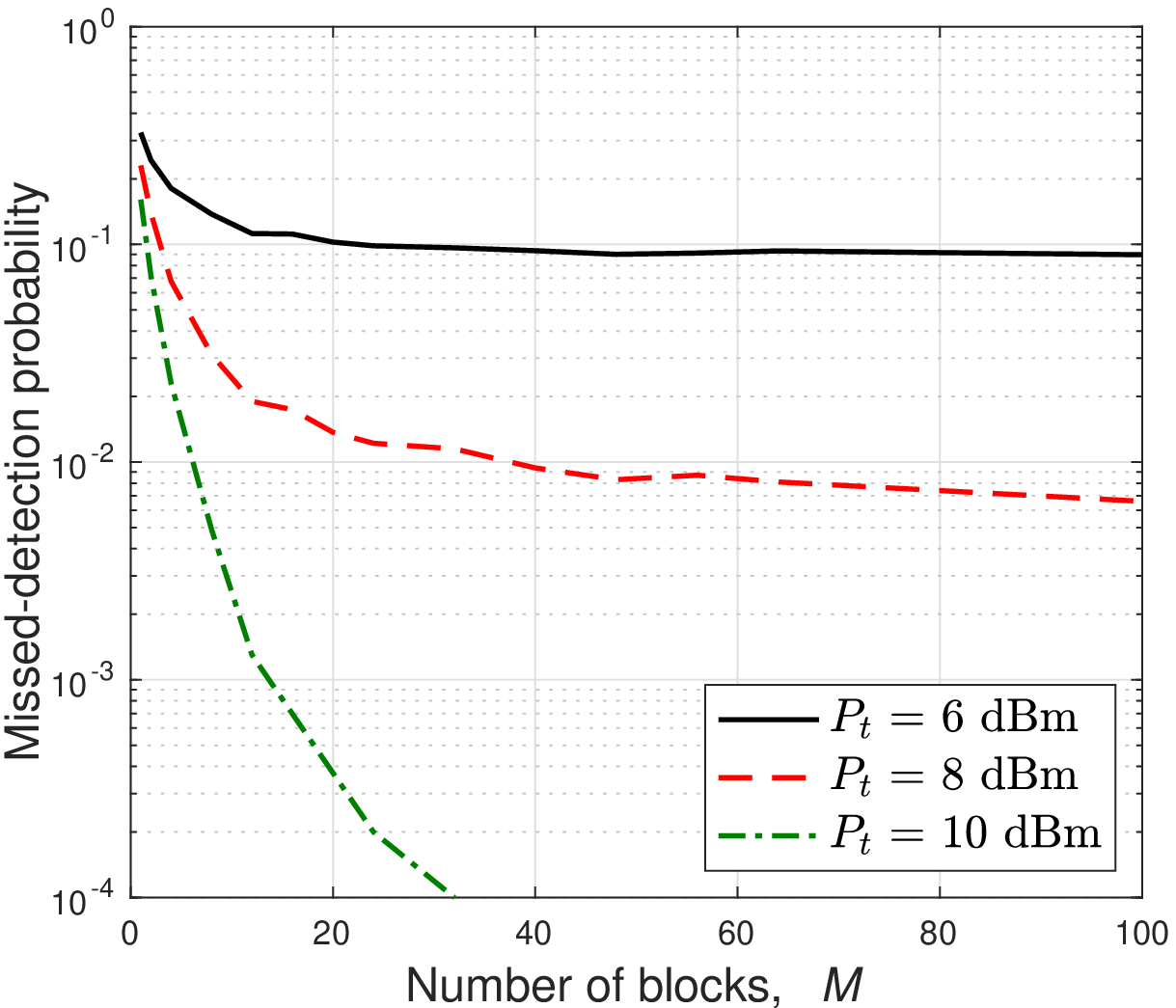}
		\label{beam}}           
	\centering    
	\hspace{0.0cm} \subfigure [Missed-detection probability versus $\alpha$ ($P_t =8$~dBm)]{
	\includegraphics[width=1.61in]{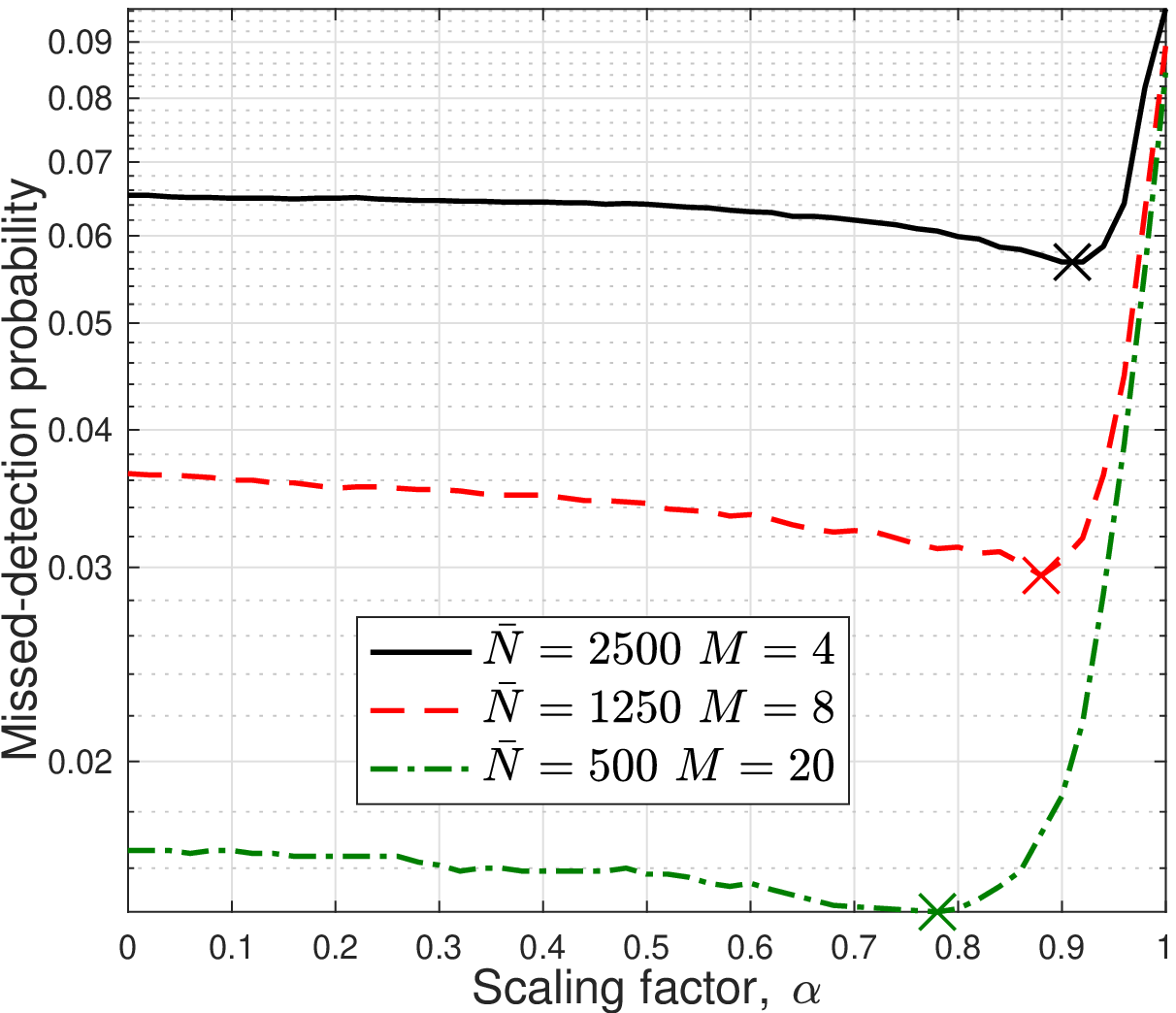} 
	\label{alpha}}     
\caption{Missed-detection probability of the proposed WED scheme versus the number of blocks $M$ or scaling factor $\alpha$, with $L=1024$ and $P_{\rm FA}^\star=0.1$. }   
\vspace*{-0.6cm}
\end{figure}

In Fig.~\ref{beam}, we show the effect of the block number $M$ on the missed-detection probability of the proposed WED scheme with $\alpha=0.5$.  
The detection threshold $\lambda$ is chosen to achieve a target false-alarm probability of $P_{\rm FA}^\star=0.1$, as required in the IEEE 802.22 standard, and $L=1024$. 
It is observed that given the total number of observations $N$, the missed-detection probability decreases as $M$ increases. 
This is because the IRS-induced  time-variant channel by changing the IRS reflection over time provides more opportunities to reap a much higher effective  channel gain as compared to that of the direct channel from the PU to SU. 
Fig.~\ref{alpha} shows the effect of the scaling factor $\alpha$ on the missed-detection probability of the proposed WED scheme, with $L=1024$, $P_{\rm FA}^\star=0.1$ and $P_t =8$~dBm. 
It is observed that there exists an optimal $\alpha$ (marked with a $\times$) which varies for different $ \bar{N}$ values. 
It is intuitive that if $ \bar{N}\to \infty$, the optimal $\alpha$ goes to $1$, since the noise effect is fully averaged out  and becomes negligible; in contrast, if $ \bar{N}\to 1$, the optimal $\alpha$ will go to $0$ to retain as many blocks as possible for averaging out the noise effect.

\section{Conclusions}\label{Conclusions}
In this letter, we considered a codebook-based IRS-aided spectrum sensing scheme, which utilizes the adjustable IRS reflection to create channel gain variations at the SU for opportunistically boosting the PU signal strength  to  facilitate its detection. 
We also  proposed a practical WED method for the spectrum sensing, which optimizes the signal power combining  coefficients for different IRS-induced channels to minimize the missed-detection probability without increasing the false-alarm probability. 
Simulation results demonstrated the substantial performance gains achieved by the proposed IRS-aided spectrum sensing scheme, as compared to various benchmark schemes.

\begin{appendices}

\appendix 
\makeatletter 
\renewcommand\theequation{A.\@arabic\c@equation } 
\makeatother 
\setcounter{equation}{0} 
	 
	 In this appendix, we show the derivation of Proposition \ref{proposition 0}.
	 By substituting \eqref{practical_combining} into \eqref{statistic} and relaxing the non-negativity restrictions, 
	 the test statistic $T$ can be reformulated as 
	 \begin{align}
	 	T = \frac{ \sum_{m=1}^{M} (T_{{m}} - \alpha   )^2 }{\sum_{ k=1}^{M}  ( T_{k} - \alpha  ) } +\alpha .
	 \end{align}
	 Moreover, by exploiting the convexity of the square function, we have the following inequality:
	 \begin{align}
	 	\left(	\frac{ \sum_{m=1}^{M} (T_{{m}} - \alpha   ) }{M}\right)^2 \le \frac{\sum_{m=1}^{M} (T_{{m}} - \alpha   )^2}{M}
	 \end{align}
	 where the equality holds if and only if $T_1=T_2=\ldots=T_M$. 
	 Therefore, $T$ is lower-bounded by
$
	 	\frac{ 1 }{M } \sum_{ m=1}^{M}  T_{m} \triangleq Z 
$,
	 and thus ${\rm Pr} \left(T_{ } > \lambda | \mathcal{H}_1 \right) $ is lower-bounded by
	 \begin{align}
	 	{\rm Pr} \left(T_{ } > \lambda | \mathcal{H}_1 \right)  \ge {\rm Pr} \left( \left. Z > \lambda \right| \mathcal{H}_1 \right)  .
	 \end{align}
	 Based on \eqref{distribution Tm appr}, it can be verified that $Z$ under $\mathcal{H}_1$ follows a Gaussian distribution with the mean of $  1 + \varrho \bar{\gamma}  $ and the variance of $  ( 1 + 2 \varrho \bar{\gamma} ) / M \bar{N} $,  where $\bar{\gamma} = \frac{1}{M}\sum_{ {m}=1}^M | g _m|^2  $. 
	 Hence, $ {\rm Pr} \left( \left. Z  > \lambda \right| \mathcal{H}_1 \right)  $ can be obtained as
	 \begin{align}
	 	{\rm Pr} \left( \left. Z  > \lambda \right| \mathcal{H}_1 \right)  = Q \left( \frac{ \lambda - 1 - \varrho \bar{\gamma}  }{ \sqrt{ ( 1 + 2\varrho \bar{\gamma} )  / M  \bar{N} } } \right)  .
	 \end{align}
	 By assuming that $M$ is sufficiently large, the exact value of $\bar{\gamma} $ can be approximated by the mean value of $| g _m|^2$, i.e., $\mathbb{E} \left[ | g _m|^2 \right] = L \beta_{PI} \beta_{IS}  + | h_{PS} |^2$ based on \eqref{gm distribution}.
	 Accordingly, the missed-detection probability $P_{\rm MD}^{ } $ is upper-bounded by
	 \begin{align}
	 	P_{\rm MD}^{ }  \le 1- Q \left( \frac{ \lambda - 1 - \varrho \left(L \beta_{PI} \beta_{IS}  + | h_{PS} |^2\right)  }{ \sqrt{ 1 + 2\varrho \left( L \beta_{PI} \beta_{IS}  + | h_{PS} |^2 \right)  } } \sqrt{ M  \bar{N} } \right)  \notag
	 \end{align}
	 thus completing the proof.

\end{appendices}

\ifCLASSOPTIONcaptionsoff
  \newpage
\fi

\bibliographystyle{IEEEtran}
\bibliography{RIS_QSM}

\end{document}